\documentclass[a4paper,11pt]{article}
\usepackage{pos}
\usepackage{booktabs}

\graphicspath{{figures/}}

\title{Looking for gravitational waves with orbital dynamics: GUEST mission and the $\mu$Hz band of Gravitational Waves}
\ShortTitle{Gravitational waves with orbital dynamics}

\author*{Diego Blas}

\affiliation{Institut de F\'isica d'Altes Energies (IFAE), The Barcelona Institute of Science and Technology,\\
  Campus UAB, 08193 Bellaterra (Barcelona), Spain\\
  and\\
  Instituci\'o Catalana de Recerca i Estudis Avan\c cats (ICREA),\\
  Passeig Llu\'is Companys 23, 08010 Barcelona, Spain}

\emailAdd{dblas@ifae.es}

\abstract{``Adding colour'' to the gravitational-wave sky by probing unexplored frequency bands has the potential to unveil new and unexpected phenomena in the Universe. The microhertz band, between pulsar timing arrays and LISA, is a particularly attractive target: it should host the extension of the signal recently found by pulsar timing arrays (if it is generated by a family of sources of gravitational waves), and it may harbour signals from the early Universe, primordial black holes or dark matter. In these proceedings, we review how the orbits of gravitationally bound systems act as resonant detectors of microhertz gravitational waves and ultra-light dark matter. We then describe GUEST (Gravitational Universe Exploration with Satellite Tracking), a mission concept recently proposed to the ESA F-class call, consisting of two dense passive spheres covered with retroreflectors, placed in highly eccentric Earth orbits and tracked by a global laser-ranging network.}

\FullConference{
4th General Meeting of COST Action COSMIC WISPers (CA21106)\\
4th Training School of the COST Action COSMIC WSIPers (CA21106) (COSMIC-WISPers2026)\\
22-25 June and 16-20 June, 2026\\
Corfu, Greece\\}

\begin{document}
\maketitle

\section{Introduction: and for gravitational waves?}
\label{sec:intro}

Humantiy has built an impressive fleet of cosmic observers that look at the Universe across the whole electromagnetic spectrum, from radiowaves of radio telescopes or the microwaves of Planck to the gamma rays of INTEGRAL. Each new window has revealed new and often unexpected phenomena. The natural question is then: \textit{and for gravitational waves?} Gravitational waves (GWs) are produced by all energetic sources with non-trivial dynamics and propagate through the primordial Universe without being scattered or absorbed. Since their first detection~\cite{LIGOScientific:2016aoc}, a new ``dark'' universe has started to manifest itself. Still, GWs have so far been directly detected only around 100~Hz, while pulsar timing arrays (PTAs) have found strong evidence for a stochastic GW background (SGWB) in the nHz band~\cite{EPTA:2023fyk,Reardon:2023gzh,NANOGrav:2023gor,Xu:2023wog}. In these proceedings we describe how to add one more colour to the GW sky by exploring the $\mu$Hz band with the dynamics of orbits. Most of the mission-related material is based on the GUEST white paper~\cite{Blas:2026xol}, where many more details can be found (see also \cite{Vetrano:2026neq}).

\section{Searches for fundamental physics in LISA and 3G detectors}
\label{sec:LISA3G}

The next generation of detectors, LISA~\cite{LISA:2024hlh} and third-generation (3G) ground-based interferometers~\cite{ET:2025xjr}, will have an extremely rich Standard Model (SM) science case: astrophysical models of black holes of several masses, of stars (white dwarfs, neutron stars...), of supernova explosions... In a sense, the community has the next 15 years clear for the SM. The situation is different for physics beyond the SM (BSM), which comprises, among other phenomena, signals from the early Universe, from dark matter, from new particles, or from modifications of gravity. In these bands, BSM signals will be superposed on a loud astrophysical foreground, and their search may require the subtraction of the astrophysical sources. How do we access new physics with GWs in quieter measurements ? A natural alternative is to \textit{move to other bands}.

\section{The landscape of gravitational-wave detectors: today to 2045}
\label{sec:landscape}

\begin{figure}[t]
\centering
\includegraphics[width=0.85\linewidth]{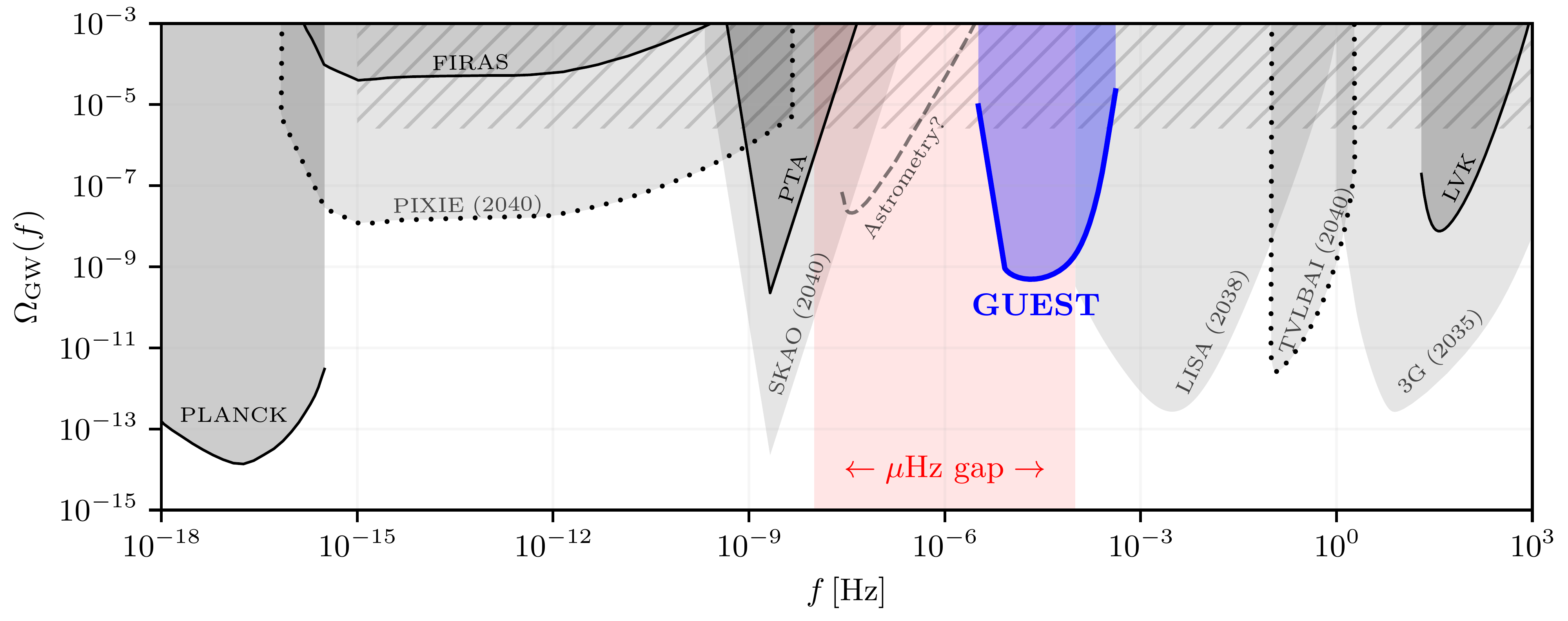}
\caption[]{The GW landscape in terms of $\Omega_{\rm GW}(f)$. Current facilities (dark grey) and planned missions (light grey) leave the $\mu$Hz band essentially uncovered. TVLBAI and PIXIE lines are estimates with no planned mission. The GUEST line corresponds to the mission concept of Sec.~\ref{sec:guest}. Adapted from~\cite{Blas:2026xol}.}
\label{fig:landscape}
\end{figure}

Detectors at very different frequencies are conveniently compared through the fractional energy density in GWs per logarithmic frequency interval,
\begin{equation}
\Omega_{\rm GW}(f)\equiv\frac{1}{\rho_c}\frac{\mathrm{d}\rho_{\rm GW}}{\mathrm{d}\log f}, \qquad \rho_{\rm GW}=\frac{c^2}{32\pi G}\langle\dot h_{ij}\dot h^{ij}\rangle,
\label{eq:Omega}
\end{equation}
where $\rho_c$ is the critical density and $h_{ij}$ the GW metric perturbation. The landscape of detectors up to about 2045 is shown in Fig.~\ref{fig:landscape}. Several regions remain essentially unexplored. Among them, the $\mu$Hz band (between PTAs and LISA, corresponding to timescales of hours to months) and the high-frequency band are particularly interesting. Here we focus on the $\mu$Hz band.

\section{The \texorpdfstring{$\mu$}{mu}Hz band}
\label{sec:muHz}

\paragraph{A guaranteed signal by the corner?} If the PTA signal~\cite{EPTA:2023fyk,Reardon:2023gzh,NANOGrav:2023gor,Xu:2023wog} is produced by supermassive black-hole binaries (SMBHBs), it must extend to the $\mu$Hz band with $\Omega_{\rm GW}\sim 10^{-8}$--$10^{-9}$~\cite{Ellis:2023owy}, making it a prime target to study the formation of massive black holes~\cite{Sesana:2019vho}. The PTA signal may also be explained by new physics (audible axions, scalar-induced GWs, phase transitions, cosmic superstrings, domain walls...)~\cite{Ellis:2023oxs}, and these interpretations predict very different spectra in the $\mu$Hz band, which makes it ideal to discriminate between them.

\begin{figure}[t]
\centering
\includegraphics[width=0.62\linewidth]{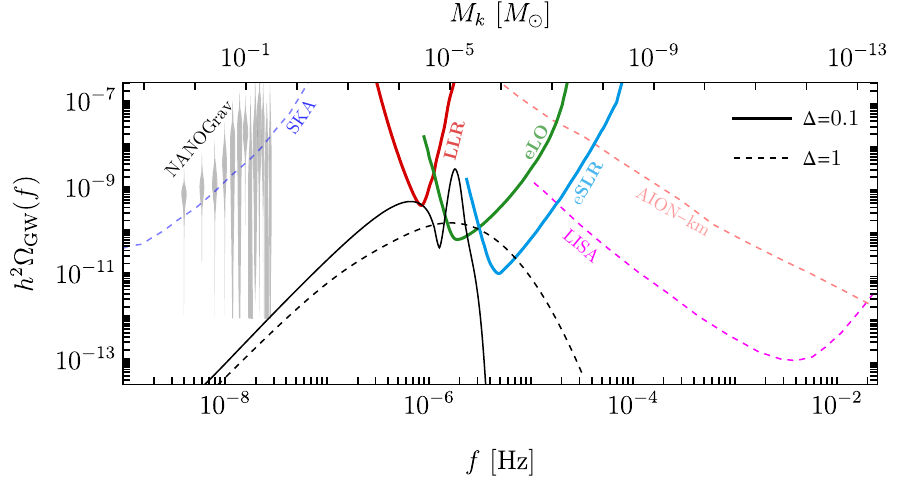}
\caption[]{Power-law-integrated sensitivity of LLR, eccentric lunar orbits (eLO) and eccentric Earth orbits tracked with SLR (eSLR)~\cite{Foster:2025nzf}, compared to NANOGrav 15-year data and future SKA, LISA and AION-km. Black lines: SIGW spectra for two benchmark PBH scenarios with widths $\Delta=0.1$ (solid) and $\Delta=1$ (dashed). From~\cite{Blas:2026xws}.}
\label{fig:sigw}
\end{figure}

\paragraph{Plentiful new physics possible.} Possible signals include features from inflation~\cite{Lasky:2015lej}, (metastable) cosmic strings~\cite{Buchmuller:2021mbb}, or first-order phase transitions (FOPTs) at $T\sim 100$~MeV~\cite{Renzini:2022alw}. PBHs are another interesting target, with GW peaks related to changes in the speed of sound during the thermal history of the Universe~\cite{Braglia:2021wwa}. A particularly clean example is that of scalar-induced GWs (SIGWs): the large curvature perturbations required to form PBHs source GWs at second order, peaking at $f\sim 10^{-9}\,{\rm Hz}\,(M_{\rm PBH}/M_\odot)^{-1/2}$~\cite{Blas:2026xws}. The $\mu$Hz band thus corresponds to $M_{\rm PBH}\sim 10^{-7}$--$10^{-4}\,M_\odot$, relevant to current microlensing candidates. As shown in Fig.~\ref{fig:sigw}, these signals can be probed with lunar laser ranging (LLR) and with satellites in eccentric orbits (eLO, eSLR). What are these lines? This is the topic of the next section.

\section{Our proposal for the \texorpdfstring{$\mu$}{mu}Hz band: binary resonance}
\label{sec:resonance}

GW interferometers track deviations of light trajectories with high precision, but these effects do not accumulate in time. On the contrary, \textit{resonant systems} target effects that do accumulate. Gravitationally bound binaries are natural resonant absorbers of GWs~\cite{Hui:2012yp}. When the GW wavelength is much larger than the binary, the relative separation $r^i$ satisfies
\begin{equation}
\ddot r^i+\frac{GM}{r^3}r^i=\frac{1}{2}\frac{\mathrm{d}^2h_{ij}}{\mathrm{d}t^2}\,r^j,
\label{eq:acc}
\end{equation}
where $M$ is the total mass and $h_{ij}$ the perturbing GW. For $h\sim\cos(2\pi f t)$ and an eccentric orbit $r\sim e\cos(2\pi t/P)$, the forcing has resonances at $f=n/P$, with $n$ an integer. For periods of a few days or less, these fall precisely in the $\mu$Hz band.

\begin{figure}[t]
\centering
\includegraphics[width=0.8\linewidth]{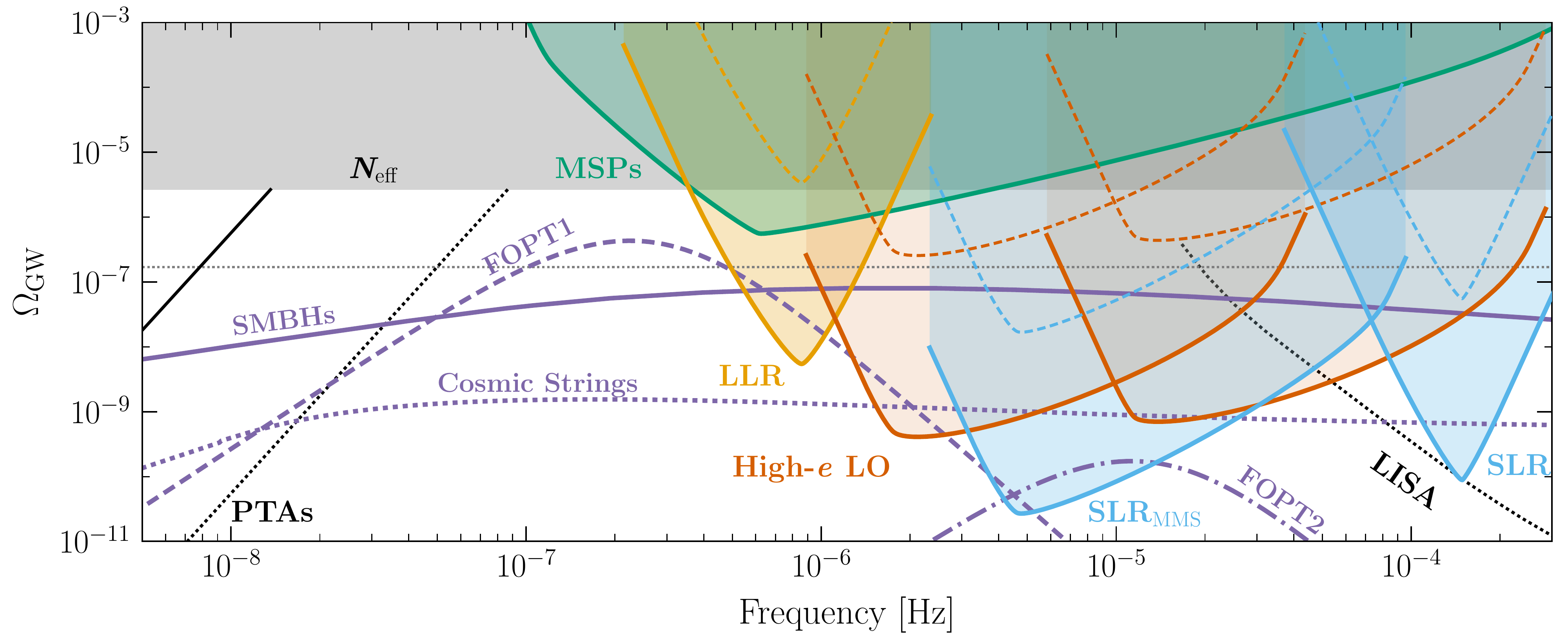}
\caption[]{Our 2025 forecasts for $\Omega_{\rm GW}$ from binary systems~\cite{Foster:2025nzf,Foster:2025csl}: the Moon (LLR), millisecond pulsars (MSPs), existing satellites (SLR LAGEOS), and proposed satellites around the Moon (High-$e$ LO) and the Earth (SLR$_{\rm MMS}$, assuming an MMS-like orbit tracked as well as LAGEOS and an idealised analysis). Dashed lines: pessimistic case with large parameter degeneracies. Possible signals from SMBHBs~\cite{Ellis:2023owy}, cosmic strings and FOPTs are also shown.}
\label{fig:2025}
\end{figure}

Two probes exploit this effect through the precise timing of binaries: binary millisecond pulsars~\cite{Jenkins:2026cjf}, and the Earth--Moon and Earth--satellite systems tracked with lunar and satellite laser ranging (LLR and SLR). Our first estimates~\cite{Blas:2021mqw,Blas:2021mpc} were based on the secular evolution of the orbital period. Going beyond the secular approximation, working directly with the observables and using \textit{all the information available}, \textit{the sensitivity grows dramatically}~\cite{Foster:2025nzf,Foster:2025csl}, partly because shifts in the true anomaly grow as $t^2$ (vs.\ $t$ for the period). The resulting forecasts for 2035 are shown in Fig.~\ref{fig:2025}: almost two orders of magnitude better than our previous estimates. The Moon already provides a remarkable sensitivity, while existing satellites such as LAGEOS, in nearly circular orbits, are mostly sensitive in the LISA band. The best $\mu$Hz sensitivity is reached by a satellite in a \textit{highly eccentric orbit with a long period} (SLR$_{\rm MMS}$ line). This line motivated GUEST.

\section{GUEST: an ESA F-class mission proposal}
\label{sec:guest}

GUEST (Gravitational Universe Exploration with Satellite Tracking) was proposed to the ESA F-class call, with the Step-2 proposal submitted in April 2026~\cite{Blas:2026xol}, by a core team from Spain, Belgium, France, Germany, Italy, Romania, the UK and the USA. The concept is simple: two dense, passive spheres covered with cube-corner retroreflectors (CCRs), in highly eccentric Earth orbits, tracked by the global SLR network of the International Laser Ranging Service (ILRS) for at least 10 years. No new technology is needed in space; the main challenge is to track satellites much farther away than usual.

\paragraph{Payload and visibility.} The spheres have an area-to-mass ratio similar to LAGEOS to suppress non-gravitational accelerations. Two designs are considered: 86~cm (LARES-2 heritage) and 70~cm (LAGEOS-like). Since the velocity changes from 1.4~km/s at apogee to 9.5~km/s at perigee, the spheres carry 1-inch CCRs optimised for perigee and 2-inch CCRs for apogee (561/189 and 366/124 units for the two designs), giving an optical cross section of 1.6--48.6 and 1.0--31.8~Mm$^2$, respectively. Current high-performance SLR stations can only track GUEST near perigee, while LLR stations can track it along the whole orbit; a 50~mJ laser upgrade would extend the reach of SLR stations to most of the orbit. Combining both, the target of 10 normal points (NPs) per orbit and satellite is realistic, and the ILRS has formally supported tracking GUEST.

\paragraph{Orbits.} The orbits (Table~\ref{tab:orbits}) were optimised for science (large period and eccentricity), end of life (natural re-entry after the 10-year mission plus a 20-year extended phase, driven by lunisolar Lidov--Kozai dynamics) and visibility from the ground stations (apogees over the Northern hemisphere, where LLR stations are). The two orbits have different orientations, which decorrelates their response and gives nearly uniform sky coverage.

\begin{table}[t]
\centering
\begin{tabular}{lcccccccc}
\toprule
Orbits at $t_0$ & Perigee alt. & Apogee alt. & AOP & RAAN & $e_0$ & $\iota$ & Period \\
\midrule
Orb1 / Orb2 & 6926 km & 86700 km & $270^\circ$ & $0^\circ$/$220^\circ$ & 0.75 & $75^\circ$/$50^\circ$ & 33.8 h \\
\bottomrule
\end{tabular}
\caption[]{Initial orbital parameters of the two GUEST satellites~\cite{Blas:2026xol}.}
\label{tab:orbits}
\end{table}

\section{Sensitivity}
\label{sec:sensitivity}

The data are two-way laser travel times. In a Keplerian model (ignoring the station position) the observable is
\begin{equation}
\Delta_\parallel=\frac{2p}{c\,(1+e\cos\theta)}+\Delta_\parallel^{n},
\label{eq:range}
\end{equation}
with $p$ the semi-latus rectum, $\theta$ the true anomaly and $\Delta_\parallel^{n}$ the noise. A resonant GW produces a modulation of $\Delta_\parallel$ with an envelope growing as $t^2$. The actual analysis simulates the orbits with all relevant accelerations and fits the GW signal together with all the parameters of precise orbit determination~\cite{Blas:2026xol}. A key issue is the orbit modelling. The standard \textit{arcs method}, derived for geodesy, re-fits the orbit every 7 days~\cite{Strugarek:2021arcs}, which largely removes the secular growth of the signal. A \textit{global analysis}, fitting a single orbit model to the whole data set, is likely optimistic at present but indicates what improved modelling can achieve. We assume 10~NPs per orbit with 10~cm precision for 10 years, which is conservative.

\begin{figure}[t]
\centering
\includegraphics[width=0.49\linewidth]{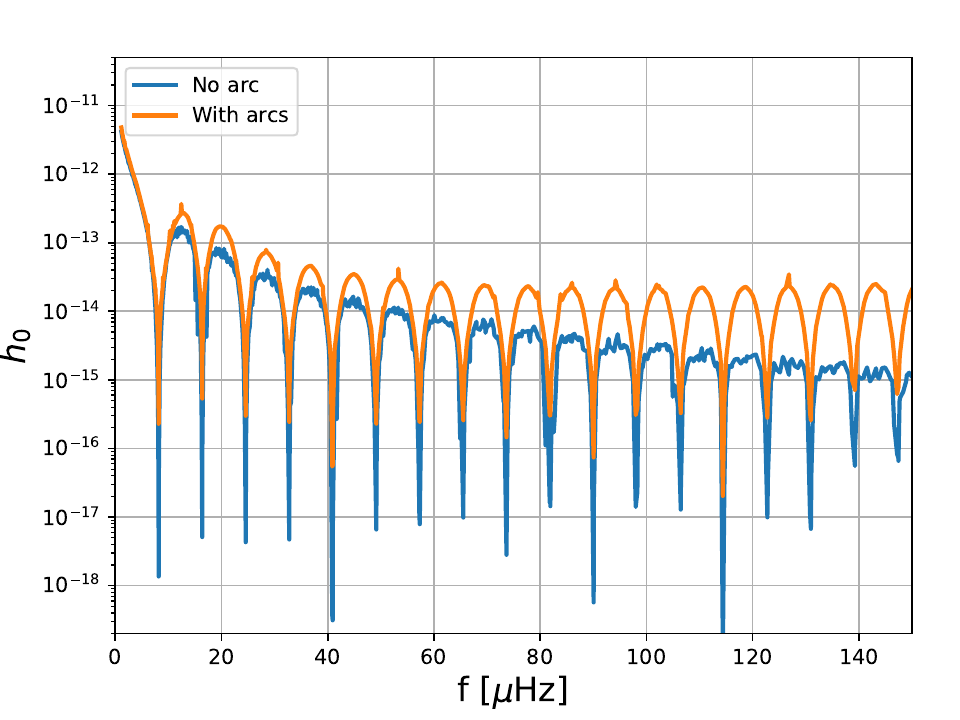}\hfill
\includegraphics[width=0.49\linewidth]{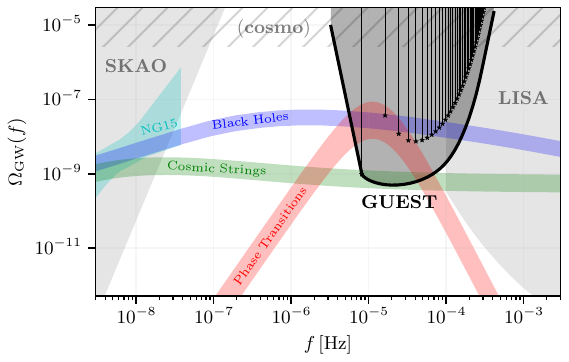}
\caption[]{GUEST 10-year sensitivity. \textbf{Left:} monochromatic GWs (SNR $=1$) for the conservative (``with arcs'') and best-case (``no arc'') analyses. \textbf{Right:} SGWB sensitivity (ideal analysis) compared to SMBHBs consistent with NANOGrav~\cite{NANOGrav:2023gor,Ellis:2023owy}, phase transitions and cosmic strings. The continuous line is the power-law integrated sensitivity~\cite{Thrane:2013oya}; the peaks are resonances. From~\cite{Blas:2026xol}.}
\label{fig:sensitivity}
\end{figure}

The results are shown in Fig.~\ref{fig:sensitivity}. For monochromatic sources, the resonant structure is clearly visible, and the two analysis strategies differ by about an order of magnitude. For a SGWB, whose effect is encoded in the covariance of the orbital elements (linear in $\Omega_{\rm GW}$), GUEST would reach $\Omega_{\rm GW}\lesssim 10^{-9}$ around $10\,\mu$Hz, probing the expected SMBHB signal and a large region of FOPT and cosmic-string models. For merging SMBHBs with $M_{\rm tot}\sim 10^8$--$10^9\,M_\odot$, the reach extends up to $z\sim 1$ in the ideal analysis, but only to a few Mpc with arcs (see also \cite{Vanvlasselaer:2026ply}): a clear example of why the data analysis must go beyond standard geodesy techniques.

\section{GUEST: more science}
\label{sec:science}

\begin{figure}[t]
\centering
\includegraphics[width=0.49\linewidth]{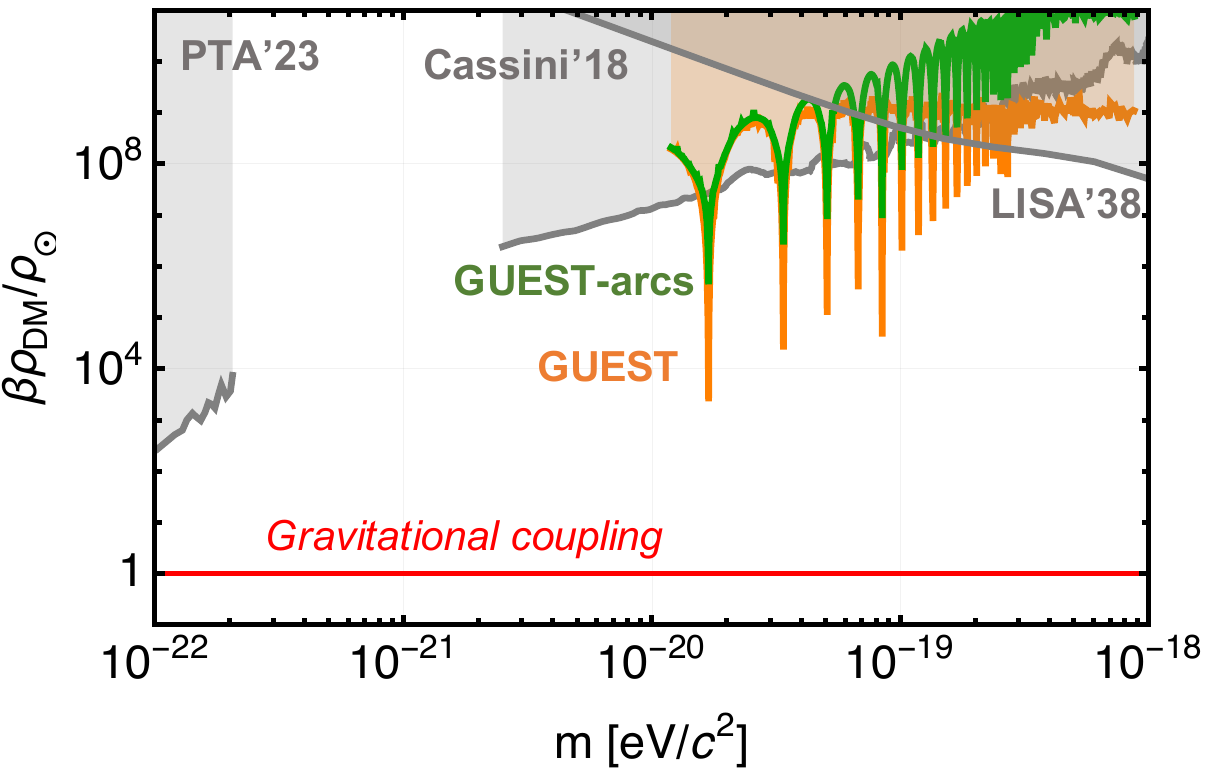}\hfill
\includegraphics[width=0.49\linewidth]{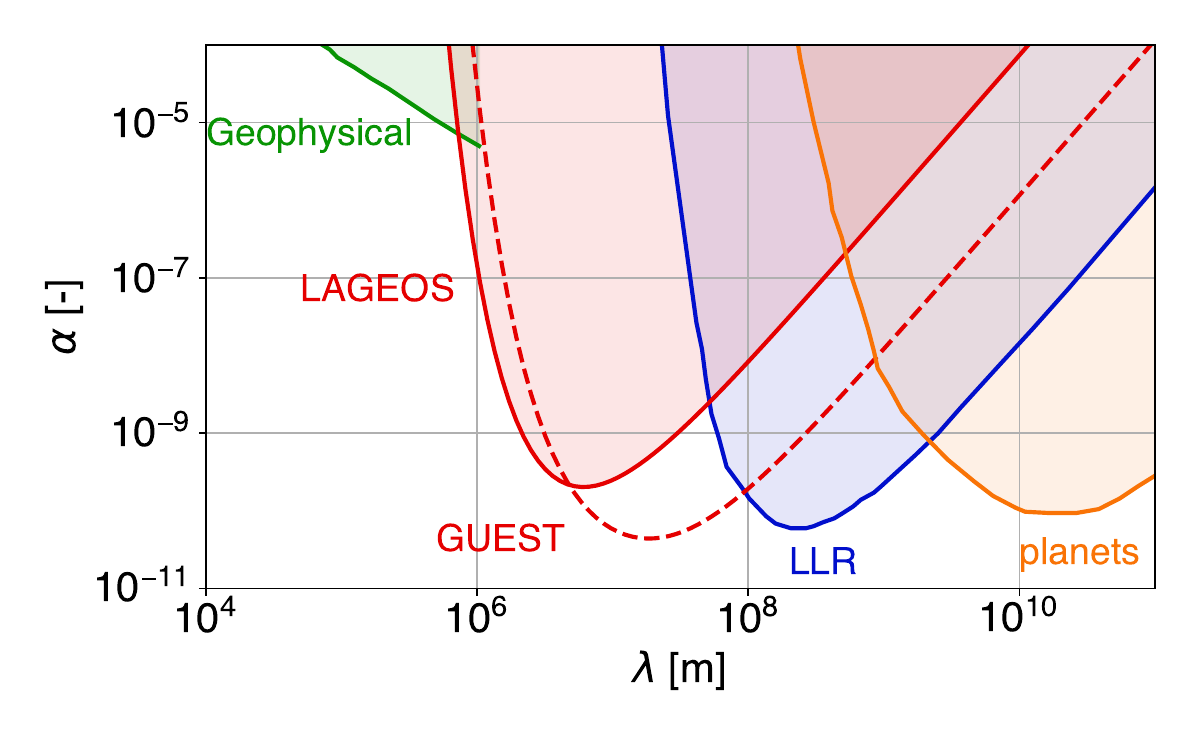}
\caption[]{\textbf{Left:} GUEST reach for ULDM in the $(m_{\rm DM},\beta\rho_{\rm DM}/\rho_\odot)$ plane, where $\beta=1$ is the universal gravitational coupling, compared to PTAs, Cassini and LISA. \textbf{Right:} constraints on a Yukawa fifth force of strength $\alpha$ and range $\lambda$, and the region opened by GUEST. From~\cite{Blas:2026xol}.}
\label{fig:more}
\end{figure}

\paragraph{Ultra-light dark matter (ULDM).} ULDM with mass $m_{\rm DM}\ll$~eV behaves as a classical field whose energy-momentum tensor makes the gravitational potential oscillate at frequency $2m_{\rm DM}$, which can resonate with binaries as GWs do~\cite{Blas:2016ddr,Blas:2019hxz}. For GUEST this corresponds to $m_{\rm DM}\sim 10^{-20}$--$10^{-19}$~eV, in a region untouched by other experiments (Fig.~\ref{fig:more}, left). Ultra-light bosons may also form superradiant clouds around nearby SMBHs (M87, NGC 3115, M31) that emit GWs in the GUEST band~\cite{Brito:2015oca}.

\paragraph{PBHs, fifth forces and geodesy.} GUEST can probe the SIGWs associated with PBHs of Sec.~\ref{sec:muHz}~\cite{Blas:2026xws}. The precision on the perigee precession scales as $1/(eP^{2/3})$, a factor $\sim 30$ better than LAGEOS, improving bounds on Yukawa fifth forces at ranges $\lambda\sim 10^{6}$--$10^{8}$~m (Fig.~\ref{fig:more}, right) and on quadratically coupled ULDM~\cite{Burrage:2026loe}. Finally, GUEST will improve the determination of $GM_\oplus$ after a few months of data, a key input for geodesy.

\section{Summary and next steps}
\label{sec:conclusions}

The $\mu$Hz band has a (nearly) guaranteed signal by the corner and plentiful possibilities for new physics. Our 2025 estimates showed that a satellite in a highly eccentric, long-period orbit is an excellent $\mu$Hz detector, which motivated GUEST: a reasonable idea for the $\mu$Hz band within the envelope of an ESA F-class or NASA SMEX mission, with a science case spanning GW astronomy, cosmology, particle physics, gravity and geodesy, and overlapping in time with LISA. Of course, there is still a long way to go. Our next steps include: building a proto-collaboration (we are where LISA was 20 years ago, but now we know a lot as a community); studying arcs vs.\ (more) global fits; realistic studies of merging SMBHBs; more physics cases and multi-band studies; decorrelation with two or more orbits; further ways to improve the orbit determination; the analysis of LAGEOS and Moon data (in progress); and data distribution and parameter determination. We hope GUEST will add some colour to the $\mu$Hz GW and dark matter sky. Be my GUEST!

\section*{Acknowledgements}

I thank the organisers of the COSMIC WISPers meeting in Corfu and all members of the GUEST collaboration, in particular A.~Bourgoin, J.~W.~Foster, A.~Hees, A.~C.~Jenkins and X.~Xue. This article is based upon work from COST Action COSMIC WISPers CA21106, supported by COST (European Cooperation in Science and Technology). 
This publication is part of the R\&D\&i project PID2023-146686NB-C31 funded by MICIU/AEI/10.13039/501100011033/ and by ERDF/EU.
IFAE is partially funded by the CERCA program of the Generalitat de Catalunya.
I acknowledge financial support from the Spanish Ministry of Science and Innovation (MICINN) through the Spanish State Research Agency, under Severo Ochoa Centres of Excellence Programme 2025-2029 (CEX2024001442-S).
This work is supported by ERC grant (GravNet, ERC-2024-SyG 101167211, DOI: 10.3030/101167211). Funded by the European Union. Views and opinions expressed are, however those of the author(s) only and do not necessarily reflect those of the European Union or the European Research Council Executive Agency. Neither the European Union nor the granting authority can be held responsible for them. 
D.Blas acknowledges the support from the European Research Area (ERA) via the UNDARK project of the Widening participation and spreading excellence programme (project number 101159929).

\end{document}